\documentclass[fleqn,usenatbib]{mnras}

\usepackage{newtxtext,newtxmath}

\usepackage[T1]{fontenc}

\DeclareRobustCommand{\VAN}[3]{#2}
\let\VANthebibliography\thebibliography
\def\thebibliography{\DeclareRobustCommand{\VAN}[3]{##3}\VANthebibliography}

\usepackage{graphicx}	
\usepackage{amsmath}	

\usepackage{siunitx}
\usepackage{ulem}
\usepackage{breqn}

\usepackage{mathrsfs}

\title[Interpebble contact radius in a comet nucleus]{Interpebble contact radius in a comet nucleus}

\author[S. Arakawa et al.]{
Sota Arakawa,$^{1}$\thanks{E-mail: arakawas@jamstec.go.jp (SA)}
Daisuke Nishiura,$^{1}$
and Mikito Furuichi$^{1}$
\\
$^{1}$Japan Agency for Marine-Earth Science and Technology, 3173-25, Showa-machi, Kanazawa-ku, Yokohama, 236-0001, Japan\\
}

\date{Accepted 2023 March 21. Received 2023 March 08; in original form 2023 January 27}

\pubyear{2023}

\begin{document}
\label{firstpage}
\pagerange{\pageref{firstpage}--\pageref{lastpage}}
\maketitle

\begin{abstract}
In recent years, the gravitational collapse of pebble clumps in the early Solar System has been regarded as a plausible scenario for the origin of comets.
In this context, ``pebbles'' represent mm- to cm-sized dust aggregates composed of (sub)micron-sized dust particles, and the structure of km-sized comets is thought to be an agglomerate of pebbles.
The contact radius for pebble--pebble contacts was modelled in an earlier study; however, the pressure dependence of the interpebble contact radius was not considered.
Here, we revisit the interpebble contact radius in a comet nucleus.
We calculated the interpebble contact radius based on JKR contact theory, and we took into consideration the effect of lithostatic pressure.
We found that the interpebble contact radius varies with depth from the surface, and the earlier model underestimated it by one order of magnitude at the centre of the comet nucleus.

\end{abstract}




\section{Introduction}

Comets are km-sized small bodies composed of volatile ices and refractory materials.
They were formed in the outer region of the Solar System, where the temperature is sufficiently lower than the sublimation temperature of ${\rm H}_{2}{\rm O}$ ice.
Their origin and evolution provide important clues about the environment of the early Solar System.

In recent years, comet formation models advocating the gravitational collapse of pebble clumps have been intensively investigated \citep[e.g.,][]{2014Icar..235..156B, 2017MNRAS.469S.755B, 2014prpl.conf..547J, 2014A&A...570A..47W, 2016A&A...587A.128L, 2021A&A...647A.126V}.
In the context of comet formation, the word ``pebbles'' means mm- to cm-sized dust aggregates composed of (sub)micron-sized dust particles originating from the interstellar medium.
In turbulent protoplanetary disks, micron-sized dust particles can grow into macroscopic pebbles via pairwise collisions \citep[e.g.,][]{2008ARA&A..46...21B, 2022ApJ...933..144A, 2022ApJ...939..100A, 2023A&A...670L..21A}.
Once the pebble radius reaches millimetres to centimetres, their collisional growth ends \citep[e.g.,][]{2010A&A...513A..57Z, 2018A&A...611A..18L}, and continued nonsticking collisions lead to rounding and compaction of pebbles \citep[e.g.,][]{2009ApJ...696.2036W, 2012Icar..218..688W}.
In gaseous protoplanetary disks, pebbles can accumulate locally via aerodynamical processes \citep[e.g.,][]{2014prpl.conf..547J, 2022arXiv220309759D}, and km-sized comets are formed by the collapse of self-gravitating pebble clumps \citep[e.g.,][]{2012Icar..221....1S}.
The structure of km-sized comets formed via gravitational collapse of pebble clumps is deemed ``hierarchical'' aggregates, i.e., agglomerates of pebbles \citep[e.g.,][]{2012Icar..221....1S, 2020MNRAS.497.1166A}.

The physical properties of pebble-pile comets have been studied extensively by using laboratory experiments \citep[e.g.,][]{2011Icar..214..286K, 2014Icar..235..156B, 2015Icar..257...33S, 2017ApJ...836...94W, 2018PhRvL.121t8001K, 2022MNRAS.514.3366M}.
\citet{2011Icar..214..286K} measured the thermal conductivity of hierarchical aggregates under high-vacuum conditions, and \citet{2012Icar..219..618G} developed a theoretical model that can reproduce the experimental results of \citet{2011Icar..214..286K}.
\citet{2012Icar..219..618G} introduced the effective surface energy of pebbles for evaluation of the interpebble contact radius, which controls heat transfer through interpebble contacts \citep[see also][]{2012Icar..218..688W}. 
\citet{2014Icar..235..156B} studied the tensile strength of hierarchical aggregates of mm-sized pebbles.
They revealed that the tensile strength of hierarchical aggregates increases with the compression pressure before breaking up.
\citet{2021MeScT..32l5301I} reported the cohesion strength between two pebbles in contact using the free-surface deformation of centrifuged hierarchical aggregates.
The cohesion strength between two pebbles increases with the interpebble contact area \citep[e.g.,][]{2020MNRAS.496.2786A}.
\citet{2015Icar..257...33S} investigated the compression curves of hierarchical aggregates, and they found that the filling factor is approximately constant when the compression pressure is lower than $10^{4}~\si{Pa}$ \citep[see also][]{2019A&A...630A...2H, 2022MNRAS.514.3366M}.
The compression behaviour of hierarchical aggregates was investigated by several studies \citep[e.g.,][]{2013Icar..226..111M, 2021PhRvR...3a3190P}.

A theoretical model of the thermal conductivity of hierarchical aggregates was developed by \citet{2012Icar..219..618G}.
This model is widely used in the thermophysical modelling of pebble-pile comets \citep[e.g.,][]{2017MNRAS.469S.685C, 2017A&A...600A.142S, 2019A&A...630A...5H, 2020MNRAS.493.3690G, 2021MNRAS.508.4705B, 2022MNRAS.514.3366M}.
In the model of \citet{2012Icar..219..618G}, the interpebble contact is modelled based on a contact theory for elastic spheres with surface energy, which is called JKR contact theory \citep{1971RSPSA.324..301J}.
Then the interpebble contact radius is a function of the effective surface energy, the elastic properties of pebbles, and the pebble radius.
We note, however, that the pressure dependence of the interpebble contact radius was not considered in \citet{2012Icar..219..618G} and their followers.
If the interpebble contact radius strongly depends on the depth from the surface, we can imagine that both the thermal and mechanical properties of the comet nucleus also vary with depth.

In this study, we revisit the interpebble contact radius in a comet nucleus.
We calculate the interpebble contact radius based on JKR contact theory, and we take into consideration the effect of lithostatic pressure.
We found that the interpebble contact radius should vary with depth from the surface.
When the effective surface energy of pebbles is equal to that of \citet{2012Icar..219..618G}, the effect of lithostatic pressure becomes important for the subsurface region where the depth is larger than $10$--$10^{2}~\si{cm}$, depending on the pebble radius.
In Section \ref{sec:model}, we review theoretical models that describe the particle--particle and pebble--pebble contacts.
In Section \ref{sec:results}, we present the results of some example calculations for a km-sized pebble-pile comet.
We briefly discuss some uncertainties and caveats for the physical properties of pebbles and pebble piles in Section \ref{sec:discussion}.
A summary is presented in Section \ref{sec:conclusions}.

\section{Model}
\label{sec:model}

\subsection{JKR contact theory \citep{1971RSPSA.324..301J}}

In this section, we review JKR contact theory for elastic spheres with surface energy and external force \citep{1971RSPSA.324..301J}.
Here, we consider the case in which two spheres in contact are made of the same material and have the same radius, $r$.
\citet{1971RSPSA.324..301J} found that the contact radius, $r_{\rm c}$, is given by 
\begin{align}
\frac{r_{\rm c}}{r} & = {\left[ \frac{3 {\left( 1 - \nu^{2} \right)}}{4 E r^{2}} {\left\{ F + \frac{3 \pi}{2} \gamma r + \sqrt{ 3 \pi \gamma r F + {\left( \frac{3 \pi}{2} \gamma r \right)}^{2} } \right\}} \right]}^{1/3}, \nonumber \\
                    & = {\left[ \frac{9 \pi \gamma {\left( 1 - \nu^{2} \right)}}{4 E r} \cdot \frac{ 1 + 2 x + \sqrt{ 1 + 2 x } }{2} \right]}^{1/3},
\label{eq:r_c}
\end{align}
where $E$ is the Young's modulus, $\nu$ is the Poisson's ratio, and $F$ is the external force acting on a sphere.
Here, $\gamma$ denotes the surface energy per unit contact area (i.e., the two surfaces in contact).
We introduce a dimensionless parameter, $x$, that describes the strength of the external force: 
\begin{equation}
x \equiv \frac{F}{3 \pi \gamma r}.
\label{eq:x}
\end{equation}

When $x \ll 1$, $r_{\rm c} / r$ becomes
\begin{equation}
\frac{r_{\rm c}}{r} \simeq {\left[ \frac{9 \pi \gamma {\left( 1 - \nu^{2} \right)}}{4 E r} \right]}^{1/3}.
\end{equation}
In contrast, when $x \gg 1$, $r_{\rm c} / r$ becomes
\begin{equation}
\frac{r_{\rm c}}{r} \simeq {\left[ \frac{3 {\left( 1 - \nu^{2} \right)} F}{4 E r^{2}} \right]}^{1/3},
\end{equation}
and the result is identical to that for Hertzian contact theory \citep{hertz1896miscellaneous}.
It should be noted that both JKR and Hertzian contact theories are applicable for $r_{\rm c} / r \ll 1$.

\subsection{Contact between cohesive elastic pebbles}

We also review the contact model for two pebbles in contact introduced by \citet{2012Icar..218..688W} and \citet{2012Icar..219..618G}.
They assumed that pebbles can be regarded as macroscopic elastic spheres with effective surface energy.
If there are no external forces acting on pebbles, the interpebble contact radius, $r_{\rm c, peb}$, is given by  
\begin{equation}
\frac{r_{\rm c, peb}}{r_{\rm peb}} = {\left[ \frac{9 \pi \gamma_{\rm peb} {\left( 1 - {\nu_{\rm peb}}^{2} \right)}}{4 E_{\rm peb} r_{\rm peb}} \right]}^{1/3},
\label{eq:r_c_peb_0}
\end{equation}
where $r_{\rm peb}$ is the pebble radius, $E_{\rm peb}$ is the effective Young's modulus of pebbles, $\nu_{\rm peb}$ is the effective Poisson's ratio of pebbles, and $\gamma_{\rm peb}$ is the effective surface energy of pebbles.

The effective surface energy of pebbles, $\gamma_{\rm peb}$, is a key parameter that controls interpebble contacts.
\citet{2012Icar..219..618G} proposed that $\gamma_{\rm peb}$ is given by the following equation \citep[see also][]{2012Icar..218..688W}\footnote{
In the original derivation by \citet{2012Icar..218..688W}, the Poisson's ratio of particles, $\nu_{\rm par}$, was used instead of $\nu_{\rm peb}$ in the calculation of $\gamma_{\rm peb}$ (Equation (\ref{eq:gamma_peb})).
We note, however, that $\nu_{\rm par} = \nu_{\rm peb} = 0.17$ is assumed and the choice of Poisson's ratio does not make any differences.
}:
\begin{align}
\gamma_{\rm peb} & = \phi_{\rm peb} {\gamma_{\rm par}}^{5/3} {\left[ \frac{9 \pi {\left( 1 - {\nu_{\rm peb}}^{2} \right)}}{r_{\rm par} E_{\rm par}} \right]}^{2/3} \nonumber \\
                 & = 5.4 \times 10^{-2}~\si{mJ.m^{-2}},
\label{eq:gamma_peb}
\end{align}
where $\phi_{\rm peb}$ denotes the filling factor inside pebbles, $\gamma_{\rm par}$ is the surface energy of particles, $r_{\rm par}$ is the particle radius, and $E_{\rm par}$ is the Young's modulus of particles.
The physical parameters used in this study are listed in Table \ref{table1} and are identical to those assumed in \citet{2020MNRAS.493.3690G} and \citet{2022MNRAS.514.3366M}.

\begin{table*}
\caption{
List of physical parameters \citep[see also][]{2020MNRAS.493.3690G}.
}
\label{table1}
\centering
\begin{tabular}{lccl}
{\bf Parameter}                      & {\bf Symbol}            & {\bf Value}                   & {\bf Reference}  \\ \hline
Particle radius                      & $r_{\rm par}$           & $1~\si{\micro m}$             & \citet{2020MNRAS.493.3690G} \\
Pebble radius                        & $r_{\rm peb}$           & $1~\si{mm}$ or $1~\si{cm}$    & \citet{2022MNRAS.514.3366M} \\
Surface energy of particles          & $\gamma_{\rm par}$      & $100~\si{mJ.m^{-2}}$          & \citet{2020MNRAS.493.3690G} \\
Surface energy of pebbles            & $\gamma_{\rm peb}$      & (see Equation (\ref{eq:gamma_peb}))             & \citet{2012Icar..219..618G} \\
Young's modulus of particles         & $E_{\rm par}$           & $55~\si{GPa}$                 & \citet{chan1973conductance} \\
Young's modulus of pebbles           & $E_{\rm peb}$           & $8.1~\si{kPa}$                & \citet{2012Icar..218..688W} \\
Poisson's ratio of particles         & $\nu_{\rm par}$         & $0.17$                        & \citet{chan1973conductance} \\
Poisson's ratio of pebbles           & $\nu_{\rm peb}$         & $0.17$                        & \citet{2012Icar..218..688W} \\
\\
Filling factor inside pebbles        & $\phi_{\rm peb}$        & $0.4$                         & \citet{2009ApJ...696.2036W} \\
Filling factor of pebble packing     & $\phi_{\rm pack}$       & $0.6$                         & \citet{2014Icar..235..156B} \\
Bulk density of the comet nucleus    & $\rho_{\rm comet}$      & $532~\si{kg.m^{-3}}$          & \citet{2016Icar..277..257J} \\
\\
Material thermal conductivity        & $k_{\rm mat}$           & (see Equations (\ref{eq:k_mat_cry}) and (\ref{eq:k_mat_amo}))     & \citet{2022MNRAS.514.3366M} \\
Rosseland mean opacity of particles  & $\kappa_{\rm R}$        & (see Equation (\ref{eq:kappa_R}))             & \citet{1994ApJ...421..640N} \\ \hline
\end{tabular}
\end{table*}

The effective Young's modulus of pebbles, $E_{\rm peb}$, is also an important parameter that controls interpebble contacts.
Based on the investigation by \citet{2012Icar..218..688W}, many studies on pebble-pile comets \citep[e.g.,][]{2017MNRAS.469S.755B, 2019A&A...630A...5H, 2020MNRAS.493.3690G, 2021MNRAS.508.4705B, 2022MNRAS.514.3366M} assumed the following value for $E_{\rm peb}$:
\begin{equation}
E_{\rm peb} = 8.1~\si{kPa}.
\label{eq:E_peb}
\end{equation}
We note that $E_{\rm peb}$ must depend on the physical properties of pebbles and constituent particles (e.g., $\phi_{\rm peb}$, $r_{\rm par}$, and $\gamma_{\rm par}$) in reality (see Section \ref{sec:E_peb}).

When external forces act on pebbles, $r_{\rm c, peb}$ should deviate from that predicted by Equation (\ref{eq:r_c_peb_0}).
Here we assumed that the dependence of $r_{\rm c, peb}$ on the external force acting on pebbles, $F_{\rm peb}$, is identical to that of Equation (\ref{eq:r_c}): 
\begin{equation}
\frac{r_{\rm c, peb}}{r_{\rm peb}} = {\left[ \frac{9 \pi \gamma_{\rm peb} {\left( 1 - {\nu_{\rm peb}}^{2} \right)}}{4 E_{\rm peb} r_{\rm peb}} \cdot \frac{ 1 + 2 x_{\rm peb} + \sqrt{ 1 + 2 x_{\rm peb} } }{2} \right]}^{1/3},
\label{eq:r_c_peb}
\end{equation}
where
\begin{equation}
x_{\rm peb} = \frac{F_{\rm peb}}{3 \pi \gamma_{\rm peb} r_{\rm peb}}.
\label{eq:x_peb}
\end{equation}

\subsection{Interpebble force}

When a pebble-pile layer is made of equal-sized spherical pebbles, the external force acting on a pebble, $F_{\rm peb}$, should be proportional to the compressive stress, $\sigma$, as follows \citep[see][]{2017AIPA....7a5310S}:
\begin{equation}
F_{\rm peb} = \frac{2 \pi {r_{\rm peb}}^{2}}{\sqrt{6} \phi_{\rm pack}} \sigma,
\label{eq:F_peb}
\end{equation}
where $\phi_{\rm pack}$ is the filling factor of pebble packing.

In this study, we assume that the compressive stress inside the comet nucleus is caused by the lithostatic pressure.
The lithostatic pressure inside a homogeneous object, $\sigma$, is given as a function of the depth $d$ as follows \citep[see][]{2017MNRAS.469S.755B}:
\begin{equation}
\sigma = \frac{2 \pi}{3} {\rho_{\rm comet}}^{2} \mathcal{G} {\left[ {R_{\rm comet}}^{2} - {\left( {R_{\rm comet}} - d \right)}^{2} \right]},
\end{equation}
where $\mathcal{G}$ is the gravitational constant, $\rho_{\rm comet}$ is the bulk density of the comet nucleus, and $R_{\rm comet}$ is the radius of the comet nucleus.
The pressure at the centre, $\sigma_{\rm centre}$, is given by
\begin{equation}
\sigma_{\rm centre} = 40 {\left( \frac{R_{\rm comet}}{1~\si{km}} \right)}^{2}~\si{Pa}.
\label{eq:sigma_center}
\end{equation}
When $d \ll R_{\rm comet}$, $\sigma$ is approximately given by
\begin{align}
\sigma & \simeq \frac{4 \pi}{3} {\rho_{\rm comet}}^{2} \mathcal{G} R_{\rm comet} d, \nonumber \\
       & \simeq 7.9 \times 10^{-2} {\left( \frac{R_{\rm comet}}{1~\si{km}} \right)} {\left( \frac{d}{1~\si{m}} \right)}~\si{Pa},
\end{align}
and $\sigma$ is proportional to both $R_{\rm comet}$ and $d$.

It is usually assumed that both $\phi_{\rm peb}$ and $\phi_{\rm pack}$ are approximately constant in a nucleus of a km-sized pebble-pile comet \citep[e.g.,][]{2017MNRAS.469S.755B, 2022Univ....8..381B}.
Here we briefly check the validity of this assumption.
When pebbles have similar radii, \citet{2017MNRAS.469S.755B} noted that $\phi_{\rm pack}$ would fall between random loose and close packing values ($\simeq 0.56$ and $\simeq0.64$, respectively).
Then $\phi_{\rm pack}$ could be nearly constant unless pebbles deform by compression, and $\phi_{\rm peb}$ barely changes when pebbles behave as elastic spheres.
As discussed in Section \ref{sec:plastic}, pebbles could indeed behave as elastic spheres and they may avoid significant deformation in the nucleus of a km-sized comet.

\section{Results}
\label{sec:results}

In this section, we present the results of some example calculations for a km-sized pebble-pile comet.
We set $R_{\rm comet} = 1~\si{km}$ in this study.
The interpebble contact radius and its dependence on depth are shown in Section \ref{sec:contact}, and the thermal conductivity in the comet nucleus is discussed in Section \ref{sec:thermal}.

\subsection{Interpebble contact radius}
\label{sec:contact}

\begin{figure}
\begin{center}
\includegraphics[width=\columnwidth]{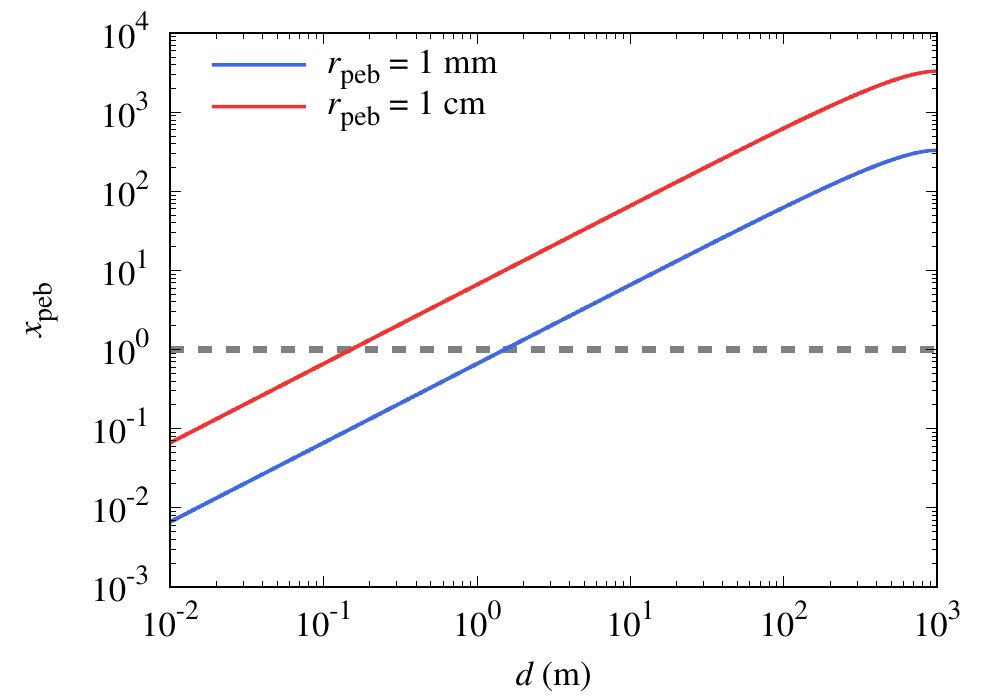}
\end{center}
\caption{
Dependence of $x_{\rm peb}$ on the depth from the surface, $d$.
The grey dashed line represents $x_{\rm peb} = 1$.
}
\label{fig:x_peb}
\end{figure}

Figure \ref{fig:x_peb} shows the dependence of $x_{\rm peb}$ on $d$.
From Equations (\ref{eq:x_peb}) and (\ref{eq:F_peb}), the dependence of $x_{\rm peb}$ on $\sigma$ and $r_{\rm peb}$ is given as follows:
\begin{equation}
x_{\rm peb} = 84 {\left( \frac{\sigma}{1~\si{Pa}} \right)} {\left( \frac{r_{\rm peb}}{1~\si{cm}} \right)}.
\end{equation}
At the centre of the comet nucleus, $x_{\rm peb}$ is orders of magnitude larger than $1$.
In contrast, we assume that $\sigma$ is given by the lithostatic pressure and $x_{\rm peb} = 0$ at the surface of the comet nucleus.

\begin{figure}
\begin{center}
\includegraphics[width=\columnwidth]{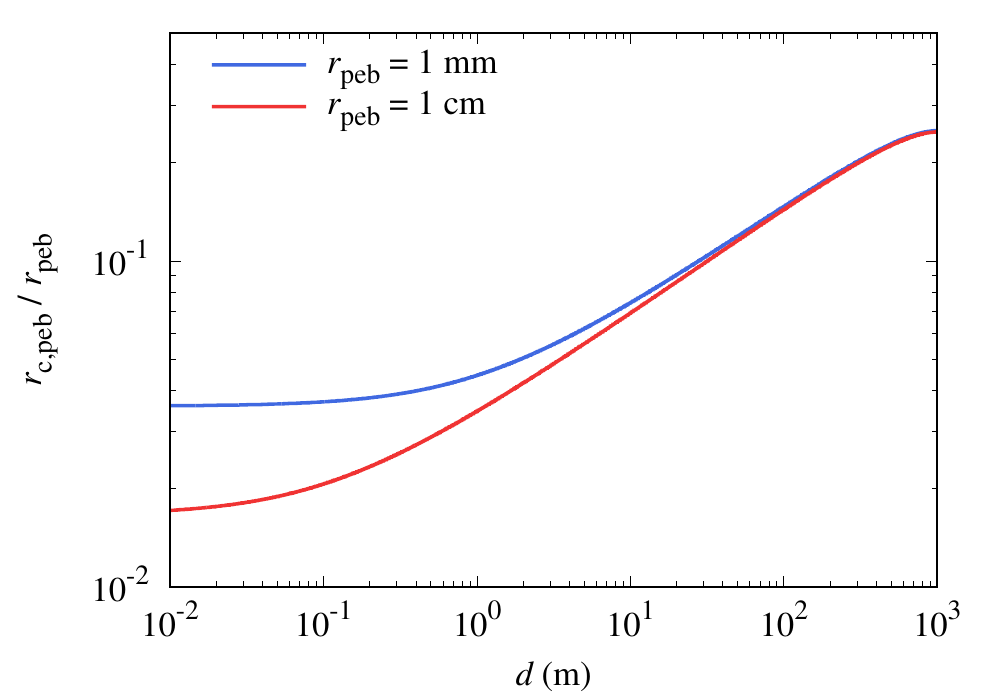}
\end{center}
\caption{
Dependence of $r_{\rm c, peb} / r_{\rm peb}$ on the depth from the surface, $d$.
}
\label{fig:r_c}
\end{figure}

For $x_{\rm peb} \ll 1$, the interpebble contact radius is given by Equation (\ref{eq:r_c_peb_0}), and $r_{\rm c, peb} / r_{\rm peb}$ is proportional to ${r_{\rm peb}}^{- 1/3}$.
In contrast, for $x_{\rm peb} \gg 1$, the interpebble contact radius becomes
\begin{equation}
\frac{r_{\rm c, peb}}{r_{\rm peb}} \simeq {\left[ \frac{\sqrt{6} \pi {\left( 1 - {\nu_{\rm peb}}^{2} \right)}}{4 \phi_{\rm pack} E_{\rm peb}} \sigma \right]}^{1/3},
\label{eq:r_c_peb_H}
\end{equation}
and we found that $r_{\rm c, peb} / r_{\rm peb}$ is nearly independent of $r_{\rm peb}$.
Figure \ref{fig:r_c} shows the dependence of $r_{\rm c, peb} / r_{\rm peb}$ on $d$.
As predicted in Equation (\ref{eq:r_c_peb_H}), $r_{\rm c, peb} / r_{\rm peb}$ is nearly independent of $r_{\rm peb}$ in the deep interior of the comet, where $x_{\rm peb} \gg 1$.
In contrast, $r_{\rm c, peb} / r_{\rm peb}$ clearly depends on $r_{\rm peb}$ but is independent of $d$ in the shallow region.

When $x_{\rm peb} \gg 1$, $r_{\rm c, peb} / r_{\rm peb}$ is given by Equation (\ref{eq:r_c_peb_H}), and it is approximated as
\begin{equation}
\frac{r_{\rm c, peb}}{r_{\rm peb}} \sim {\left( \frac{\sigma}{E_{\rm peb}} \right)}^{1/3}.
\end{equation}
As $\sigma_{\rm centre}$ is two orders of magnitude lower than $E_{\rm peb}$, $r_{\rm c, peb} / r_{\rm peb} \ll 1$ is satisfied in the entire comet nucleus.
Thus the JKR contact theory (Equation (\ref{eq:r_c_peb})) could be applicable for interpebble contacts in comet nuclei if pebbles behave as elastic spheres.

In Figure \ref{fig:x_peb}, we see that $x_{\rm peb} < 1$ in the subsurface region of $d \lesssim 1~\si{m}$ for $r_{\rm peb} = 1~\si{mm}$.
Even for $r_{\rm peb} = 1~\si{cm}$, the subsurface region of $d \lesssim 10~\si{cm}$ satisfies $x_{\rm peb} < 1$.
For comet 67P/Churyumov--Gerasimenko, which is the target of the {\it Rosetta} mission, the diurnal and seasonal thermal skin depths are evaluated as $\delta_{\rm diurnal} \sim 1~\si{cm}$ and $\delta_{\rm seasonal} \sim 1~\si{m}$, respectively \citep[e.g.,][]{2020MNRAS.497.1166A, 2020SSRv..216...44C}.
Our results indicate that the diurnal variation in the subsurface temperature reflects the thermophysical properties of the uncompressed layer where $x_{\rm peb} < 1$, while the seasonal variation might reflect the thermophysical properties of the compressed region where $x_{\rm peb} \gtrsim 1$.

\subsection{Thermal conductivity}
\label{sec:thermal}

The increase in $r_{\rm c, peb} / r_{\rm peb}$ in the comet nucleus could have large impacts on the physical properties.
Here, we discuss the impacts of the increase in $r_{\rm c, peb} / r_{\rm peb}$ on the thermal conductivity.
The thermal conductivity of a pebble-pile comet, $k_{\rm comet}$, is given by the sum of two terms \citep[e.g.,][]{2012Icar..219..618G, 2020MNRAS.493.3690G}:
\begin{equation}
k_{\rm comet} = k_{\rm net} + k_{\rm rad},
\end{equation}
where $k_{\rm net}$ is the thermal conductivity through the solid particle contacts (i.e., network conduction; see Appendix \ref{app:k_net}) and $k_{\rm rad}$ is the thermal conductivity due to radiation through the void space between the pebbles (see Appendix \ref{app:k_rad}).
We note that $k_{\rm net}$ is proportional to $r_{\rm c, peb} / r_{\rm peb}$, while $k_{\rm rad}$ is independent of $r_{\rm c, peb} / r_{\rm peb}$.
Therefore, $k_{\rm comet}$ significantly depends on $d$ when $k_{\rm net} \gg k_{\rm rad}$, while $k_{\rm comet}$ is nearly independent of $d$ when $k_{\rm net} \ll k_{\rm rad}$.

\begin{figure}
\begin{center}
\includegraphics[width=\columnwidth]{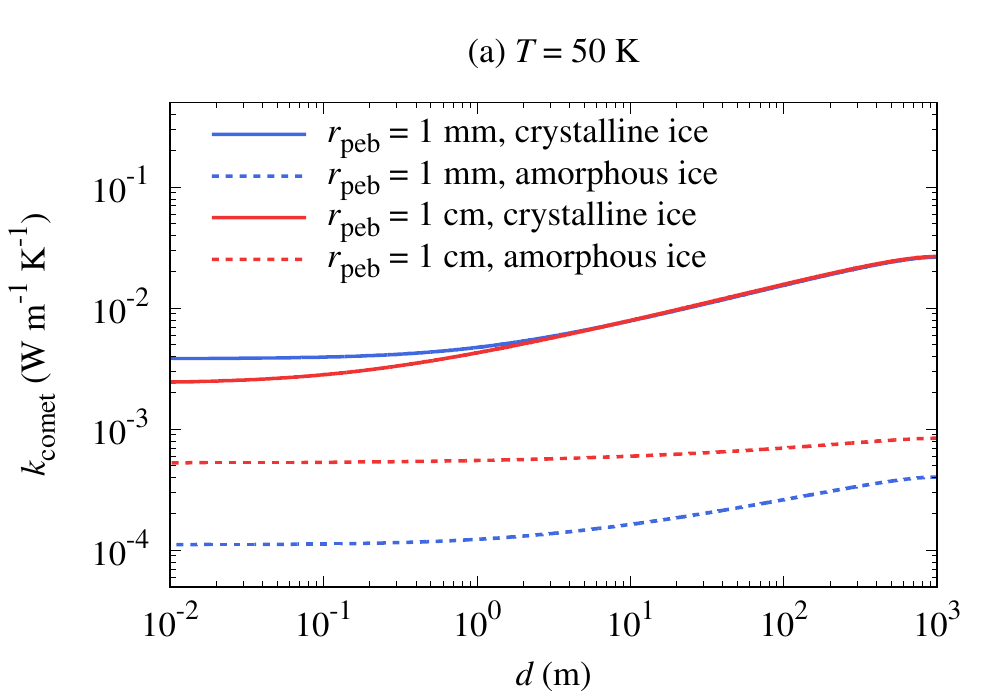}
\includegraphics[width=\columnwidth]{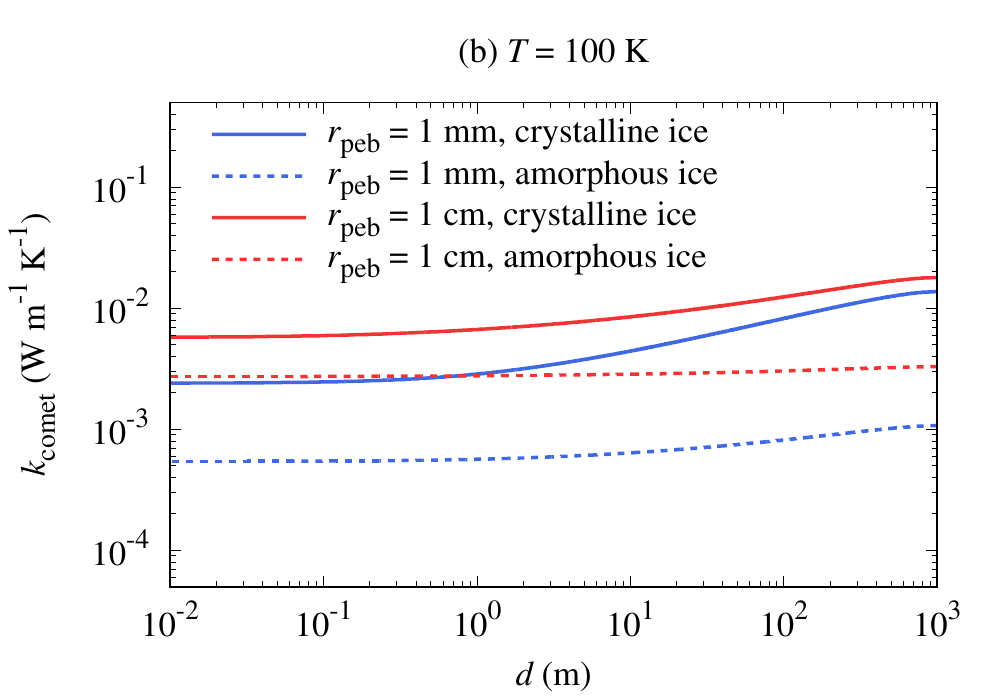}
\includegraphics[width=\columnwidth]{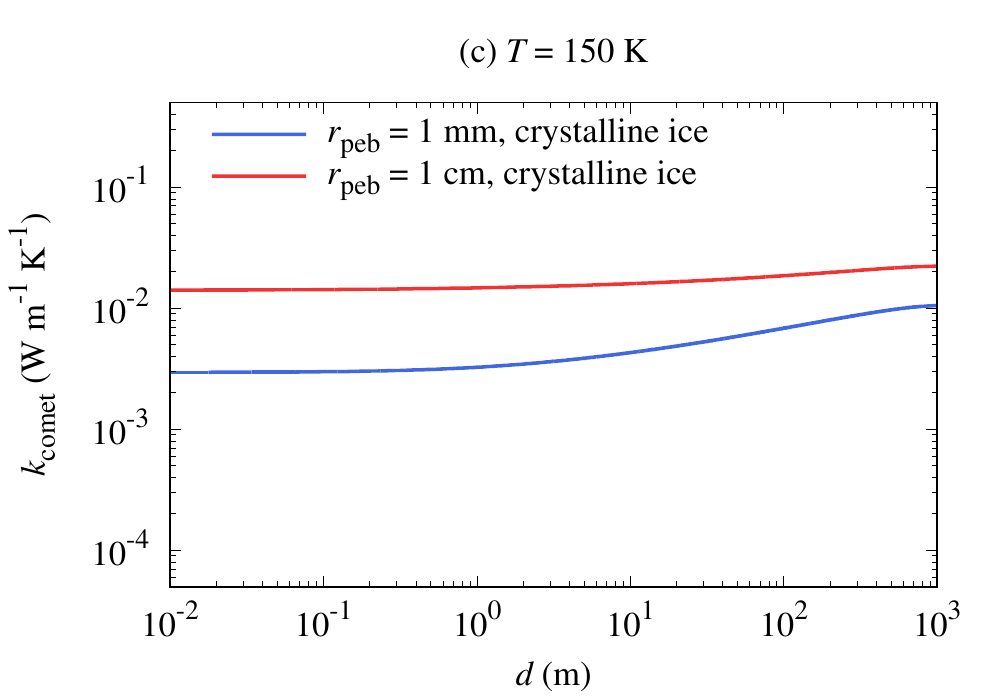}
\end{center}
\caption{
Dependence of $k_{\rm comet}$ on the depth from the surface, $d$.
(a) For $T = 50~\si{K}$.
(b) For $T = 100~\si{K}$.
(c) For $T = 150~\si{K}$.
}
\label{fig:k_comet}
\end{figure}

\begin{figure}
\begin{center}
\includegraphics[width=\columnwidth]{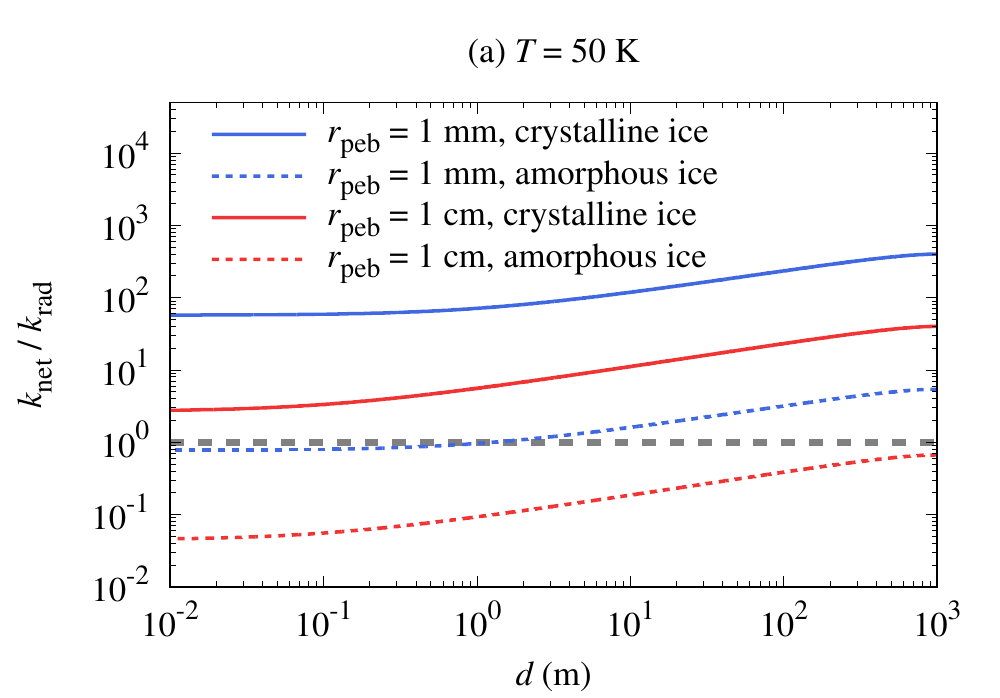}
\includegraphics[width=\columnwidth]{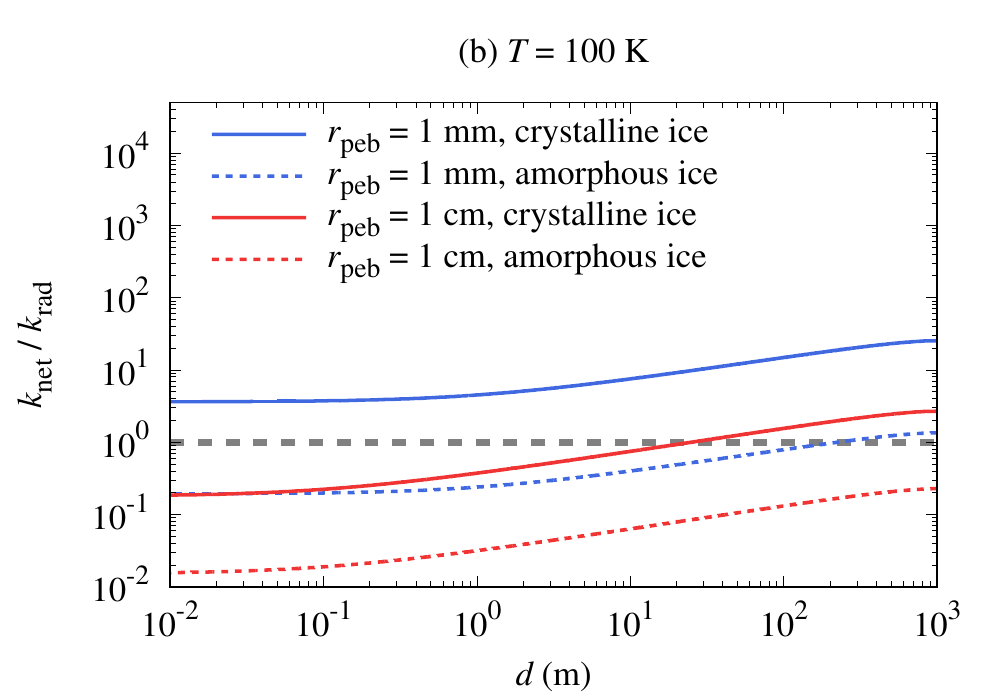}
\includegraphics[width=\columnwidth]{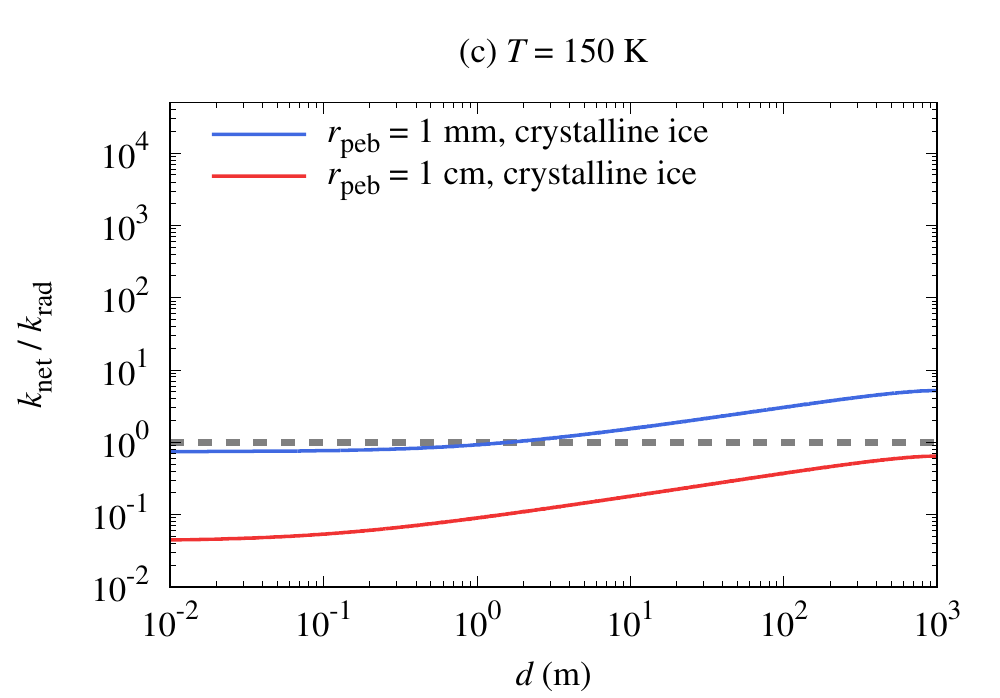}
\end{center}
\caption{
Dependence of $k_{\rm net} / k_{\rm rad}$ on the depth from the surface, $d$.
(a) For $T = 50~\si{K}$.
(b) For $T = 100~\si{K}$.
(c) For $T = 150~\si{K}$.
The grey dashed line represents $k_{\rm net} / k_{\rm rad} = 1$.
}
\label{fig:k_net_k_rad}
\end{figure}

Figure \ref{fig:k_comet} shows the dependence of $k_{\rm comet}$ on $d$, and Figure \ref{fig:k_net_k_rad} shows $k_{\rm net} / k_{\rm rad}$ as a function of $d$.
In this study, we assume that the material thermal conductivity of constituent particles, $k_{\rm mat}$, is set to be equal to that of ${\rm H}_{2}{\rm O}$ ice for simplicity.
It is known that the $k_{\rm mat}$ of ice particles strongly depends on whether they are crystalline or amorphous.
The dependence of $k_{\rm mat}$ on $T$ is given by Equation (\ref{eq:k_mat_cry}) for crystalline ice and Equation (\ref{eq:k_mat_amo}) for amorphous ice.
We also assume that the temperature is homogeneous within the comet nucleus for simplicity.

We found that $k_{\rm comet}$ at $T = 50~\si{K}$ is strongly dependent on $d$ when pebbles are made of crystalline ice (Figure \ref{fig:k_comet}(a)); $k_{\rm comet}$ at the centre is approximately an order of magnitude higher than $k_{\rm comet}$ at the surface.
Figure \ref{fig:k_net_k_rad}(a) shows that $k_{\rm net} \gg k_{\rm rad}$ is satisfied for this case.
Therefore, $k_{\rm comet}$ is approximately proportional to $r_{\rm c, peb} / r_{\rm peb}$ and increases with $d$, as shown in Figure \ref{fig:r_c}.
In contrast, $k_{\rm comet}$ at $T = 50~\si{K}$ barely depends on $d$ when the pebble radius is $r_{\rm peb} = 1~\si{cm}$ and they are made of amorphous ice.
This is reasonable because $k_{\rm net} / k_{\rm rad} \ll 1$ in this case and $k_{\rm rad}$ is independent of $d$.

The dependence of $k_{\rm comet}$ on $d$ at higher temperatures is shown in Figures \ref{fig:k_comet}(b) and \ref{fig:k_comet}(c).
For a fixed pebble radius and state of ice, $k_{\rm net} / k_{\rm rad}$ decreases with increasing $T$ (see Figure \ref{fig:k_net_k_rad}).
At $T = 150~\si{K}$, $k_{\rm comet}$ barely depends on $d$ when the pebble radius is $r_{\rm peb} = 1~\si{cm}$ and they are made of crystalline ice.
We note that amorphous ice particles will crystallize at approximately $T \sim 100~\si{K}$ \citep[e.g.,][]{2021ApJ...918...45K, 2022MNRAS.514.3366M}, and only crystalline ice is considered at $T = 150~\si{K}$.

The early thermal evolution of comets, during and immediately after formation, is the result of heating by radioactive decay of ${}^{26}{\rm Al}$ and cooling by heat transfer from the hot comet nucleus to the cold space \citep[e.g.,][]{2008SSRv..138..147P, 2017ApJ...842...11S, 2022MNRAS.514.3366M}.
The cooling rate is proportional to the thermal conductivity, and the heating rate is proportional to the abundance of ${}^{26}{\rm Al}$, which decreases exponentially with time.
The increase in $k_{\rm comet}$ in the comet nucleus should have impacts on the heating and cooling history of comets.

\section{Discussion}
\label{sec:discussion}

\subsection{Possible impacts of an increase in the interpebble contact radius on the strength of the comet nucleus}

In Section \ref{sec:thermal}, we discuss the impact of an increase in the interpebble contact radius on the thermal conductivity.
However, not only the thermal properties but also the strengths of pebble piles should depend on the interpebble contact radius.
Here we briefly discuss possible impacts on the strength of the comet nucleus.

We revealed that the interpebble contact radius increases with depth, and $r_{\rm c, peb} / r_{\rm peb}$ at the centre of comets could be an order of magnitude larger than that at the surface (Figure \ref{fig:r_c}).
The interpebble contact area also increases with depth by two orders of magnitude.
This may cause an increase in cohesion between pebbles, and the tensile and shear strengths of pebble piles might be depth-dependent.
Indeed, laboratory experiments by \citet{2014Icar..235..156B} reported that the tensile strength of pebble piles increases with the compression pressure before breaking up.
We will discuss the possible impacts of depth-dependent strengths on the geological evolution of pebble-pile comets in future studies.

\subsection{Model caveats}
\label{sec:caveats}

In this study, we calculated the interpebble contact radius and its dependence on depth using the material properties of pebbles (e.g., $\gamma_{\rm peb}$ and $E_{\rm peb}$) derived by earlier studies \citep{2012Icar..219..618G, 2012Icar..218..688W}.
These parameters for pebbles are significantly different from those for constituent particles (see Table \ref{table1}), and our results shown in Section \ref{sec:results} strongly depend on both $\gamma_{\rm peb}$ and $E_{\rm peb}$.
Here we discuss some uncertainties and caveats for these physical properties of pebbles.

\subsubsection{Effective surface energy of pebbles}

Although the canonical model of the effective surface energy of pebbles (Equation (\ref{eq:gamma_peb})), which is derived by \citet{2012Icar..219..618G}, is widely used in earlier studies, a potential problem was noted by \citet{2020MNRAS.496.2786A}.
\citet{2016A&A...593A...3B} investigated the critical pull-off force for separating two sticking pebbles, $F_{\rm po}$, and they found that $F_{\rm po} \sim 10^{-7}~\si{N}$ when the pebble radius is $r_{\rm peb} \sim 0.1~\si{mm}$ and constituent particles are micron-sized ${\rm Si}{\rm O}_{2}$ spheres.
\citet{2020MNRAS.496.2786A} revealed that the value of $F_{\rm po}$ obtained from laboratory experiments \citep{2016A&A...593A...3B} is an order of magnitude larger than that calculated from the effective surface energy modelled by \citet{2012Icar..219..618G}.
In other words, Equation (\ref{eq:gamma_peb}) might underestimate the value of $\gamma_{\rm peb}$ when we discuss the critical pull-off force for separating two sticking pebbles.

For $x_{\rm peb} \gg 1$, $r_{\rm c, peb} / r_{\rm peb}$ is nearly independent of $\gamma_{\rm peb}$ (Equation (\ref{eq:r_c_peb_H})).
In contrast, $r_{\rm c, peb} / r_{\rm peb}$ increases with $\gamma_{\rm peb}$ when $x_{\rm peb} \ll 1$, $r_{\rm c, peb} / r_{\rm peb}$ (Equation (\ref{eq:r_c_peb_0})).
Therefore, the thermal and physical properties of pebble piles in the shallow region ($d \ll 1~\si{m}$) would be strongly affected by the effective surface energy of pebbles, while the thermal evolution of the comet nucleus may not be sensitive to $\gamma_{\rm peb}$.

We note that $\gamma_{\rm peb}$ must depend on the surface composition of the constituent particles.
In earlier studies on the thermophysical modelling of comets \citep[e.g.,][]{2020MNRAS.493.3690G, 2022MNRAS.514.3366M}, a constant value of $\gamma_{\rm par} = 100~\si{mJ.m^{-2}}$ is assumed not only for pebbles composed of refractory particles but also for pebbles composed of icy particles.
As Equation (\ref{eq:gamma_peb}) predicted that $\gamma_{\rm peb}$ is proportional to ${\gamma_{\rm par}}^{5/3}$, the dependence of $\gamma_{\rm peb}$ on the surface composition of constituent particles would be of great importance.

The composition of ice in the outer Solar System has been extensively studied \citep[e.g.,][]{2019ARA&A..57..113A, 2020ApJ...899..134K, 2021PhR...893....1O}. 
Except for ${\rm H}_{2}{\rm O}$, ${\rm C}{\rm O}$ and ${\rm C}{\rm O}_{2}$ are thought to be the most prominent species in the outer Solar System, and laboratory experiments of collisions between ${\rm C}{\rm O}_{2}$ ice aggregates composed of micron-sized particles were performed by \citet{2021ApJ...923..134F}.
It is known that the material properties ($\gamma_{\rm par}$, $E_{\rm par}$, and $\nu_{\rm par}$) of ${\rm C}{\rm O}_{2}$ ice are close to those of ${\rm H}_{2}{\rm O}$ ice \citep[e.g.,][]{2021ApJ...910..130A, 2022MNRAS.512.3754F}, while the material properties of ${\rm C}{\rm O}$ ice are poorly understood.
The properties of icy particles may also depend on whether refractory materials are mantled by the icy layer \citep[see Figure 1 of][]{2020MNRAS.497.1166A} or whether refractory and icy materials are mixed well within each particle.
Future studies on these points are essential to develop the physical modelling of pebble-pile comets.

\subsubsection{Effective Young's modulus of pebbles}
\label{sec:E_peb}

The canonical value of the effective Young's modulus of pebbles ($E_{\rm peb} = 8.1~\si{kPa}$; see Equation (\ref{eq:E_peb})) is also widely used in earlier studies on pebble-pile comets.
\citet{2012Icar..218..688W} performed microgravity experiments of free collisions between pebbles, and they found that the sticking velocity is $v_{\rm stick} = 2.1 \times 10^{-4}~\si{m.s^{-1}}$ for $r_{\rm peb} = 0.5~\si{mm}$.
Based on JKR contact theory, \citet{2012Icar..218..688W} derived the following equation for $v_{\rm stick}$ \citep[see also][]{THORNTON1998154}:
\begin{equation}
v_{\rm stick} = 4.2 {\left( \frac{{\gamma_{\rm peb}}^{5} {r_{\rm peb}}^{4}}{{m_{\rm peb}}^{3} {E_{\rm peb}}^{2}} \right)}^{1/6},
\end{equation}
where $m_{\rm peb} = {( 4 \pi / 3 )} \rho_{\rm peb} {r_{\rm peb}}^{3}$ is the pebble mass, and $\rho_{\rm peb}$ is the density of pebbles.
Then \citet{2012Icar..218..688W} reported the value of $E_{\rm peb} = 8.1~\si{kPa}$ for pebbles whose filling factor is $\phi_{\rm peb} = 0.35$ and constituent particles are micron-sized ${\rm Si}{\rm O}_{2}$ spheres.

We can imagine that $E_{\rm peb}$ must depend on many parameters including the filling factor of pebbles and the composition and radius of constituent particles, as is the case for $\gamma_{\rm peb}$.
\citet{2022MNRAS.509.5641S} investigated the collisional properties of cm-sized porous dust aggregates whose filling factor is approximately $10\%$.
Then, they developed a method to calculate the dependence of $E_{\rm peb}$ (and $\nu_{\rm peb}$) on $\phi_{\rm peb}$.
Their model indicates that $E_{\rm peb} \sim 40$--$60~\si{kPa}$ for icy pebbles when $\phi_{\rm peb} \sim 0.1$--$0.2$ and $r_{\rm par} \sim 3~\si{\micro m}$, and $E_{\rm peb} > 60~\si{kPa}$ for $\phi_{\rm peb} > 0.2$.
The model also predicts that $E_{\rm peb}$ increases with decreasing $r_{\rm par}$.
The typical particle radius in the interstellar medium is $r_{\rm par} \sim 0.1~\si{\micro m}$ \citep[e.g.,][]{1977ApJ...217..425M}.
\citet{2022A&A...663A..57T} revealed that the results of optical and near-infrared polarimetric observations are consistent with dust aggregates in extrasolar protoplanetary disks that are also composed of submicron-sized particles.
Thus, the value of $E_{\rm peb}$ for {\it natural} pebble piles (i.e., comets in the Solar System) might be orders of magnitude higher than that assumed in this study ($E_{\rm peb} = 8.1~\si{kPa}$).
We will discuss this point in the future by numerical simulations using the discrete element method \citep[e.g.,][]{2018NatSR...8.8685F}.

It is important to note that $r_{\rm c, peb} / r_{\rm peb}$ is proportional to ${E_{\rm peb}}^{- 1/3}$ for the cases of both $x_{\rm peb} < 1$ and $x_{\rm peb} > 1$ (Equation (\ref{eq:r_c_peb})).
If the true value of $E_{\rm peb}$ is three orders of magnitude larger than the canonical value, the thermal conductivity through the solid particle
contacts would be one order of magnitude smaller than shown in Section \ref{sec:thermal}.
In this case, $k_{\rm comet}$ for amorphous ice (dashed lines in Figure \ref{fig:k_comet}) is approximately given by $k_{\rm comet} \simeq k_{\rm rad}$ and is independent of $d$ when the temperature and the pebble radius are constant over depth.

\subsubsection{Plastic yielding of pebbles}
\label{sec:plastic}

In this study, we assume that pebbles behave as elastic spheres.
However, it is evident that pebbles no longer behave as elastic spheres when $F_{\rm peb}$ exceeds the threshold for plastic yielding.
\citet{2022A&A...664A.147O} measured the peak compression force at the yielding point, $F_{\rm peak}$, and they found that $F_{\rm peak} / {( \pi {r_{\rm peb}}^{2} )} = {( 18 \pm 7 )}~\si{kPa}$ for dry pebbles of $r_{\rm peb} = 1.5~\si{mm}$, $\phi_{\rm peb} = 0.26$, and constituent particles are $r_{\rm par} = 2.5~\si{\micro m}$-sized glass beads.
The value of $F_{\rm peb} / {( \pi {r_{\rm peb}}^{2} )}$ at the centre of a km-sized comet is roughly given by $\sigma_{\rm centre} = 40 {( R_{\rm comet} / 1~\si{km} )}^{2}~\si{Pa}$ (see Equation (\ref{eq:sigma_center})), which is orders of magnitude lower than $F_{\rm peak} / {( \pi {r_{\rm peb}}^{2} )}$.
Thus, dry pebbles used in \citet{2022A&A...664A.147O} might behave as elastic spheres in a comet nucleus.

The value of $F_{\rm peak} / {( \pi {r_{\rm peb}}^{2} )} = {( 18 \pm 7 )}~\si{kPa}$ may also be consistent with the compression curves of hierarchical aggregates \citep[e.g.,][]{2015Icar..257...33S, 2019A&A...630A...2H}.
They reported that the total filling factor of pebble piles is approximately constant when $\sigma < 10~\si{kPa}$.
It is thought that compression of pebble piles starts when deformation and destruction of pebbles occurs \citep[e.g.,][]{2022Univ....8..381B}.
We note that $F_{\rm peak}$ must depend on many parameters, including $\phi_{\rm peb}$ and $r_{\rm par}$.
In addition, $F_{\rm peak}$ increases significantly when pebbles are sintered \citep[e.g.,][]{2022A&A...664A.147O}.
As most related experiments are for pebbles composed of micron-sized glass beads, future studies on the onset of plastic yielding of icy pebbles are necessary.

\section{Conclusions}
\label{sec:conclusions}

The physical properties of pebble piles have been studied extensively by laboratory experiments.
These studies highlighted the importance of pebble--pebble contacts for the physical modelling of pebble-pile comets.
The contact radius for pebble--pebble contacts was modelled by \citet{2012Icar..219..618G}; however, the pressure dependence of the interpebble contact radius was not considered in their model.

In this study, we revisited the interpebble contact radius in a comet nucleus (see Section \ref{sec:model}).
We calculated the interpebble contact radius based on JKR contact theory \citep{1971RSPSA.324..301J}, and we took into consideration the effect of lithostatic pressure.
We found that the interpebble contact radius varies with depth from the surface, and the earlier model underestimated it by one order of magnitude at the centre of the comet nucleus (see Section \ref{sec:contact}).

We also evaluated the impact of the increase in the interpebble contact radius on the thermal conductivity of a pebble-pile comet (see Section \ref{sec:thermal}).
We found that when pebbles are cm-sized and made of crystalline ice, the thermal conductivity through interpebble contacts is larger than the thermal conductivity due to radiation through the void space between pebbles.
In this case, the thermal conductivity of a pebble-pile comet is proportional to the interpebble contact radius, which increases with depth.
In contrast, when pebbles are mm-sized and made of amorphous ice, the thermal conductivity of a pebble-pile comet is nearly independent of the depth.
The impacts of modification of the thermal conductivity on the early thermal evolution of comets will be discussed in our future studies.

We note that both the effective surface energy and Young's modulus of pebbles are still under debate (see Section \ref{sec:caveats}).
These parameters should depend on many physical parameters of pebbles and their constituent particles, and several prediction models have been proposed \citep[e.g.,][]{2012Icar..219..618G, 2022MNRAS.509.5641S}.
Recently, \citet{2022MNRAS.509.5641S} proposed a novel model to describe the filling factor dependence of the effective surface energy and Young's modulus of pebbles.
Their proposed model was used for interpreting their experimental results on dynamic pebble--pebble collisions.
Although their model was originally derived for the case of $\phi_{\rm peb} < 0.33$, it might be worthwhile to investigate the quasistatic behavior of pebble piles on the basis of their model in future studies.

It is also a problem that the physical properties of pebbles composed of submicron-sized particles are barely investigated in laboratories, although comets in the Solar System would be made of pebbles composed of submicron-sized particles.
As submicron-sized particles are usually used in numerical simulations, future collaborations between numerical simulations and laboratory experiments might be the key to unveiling the nature of pebble-pile comets.

\section*{Acknowledgements}

The authors wish to express their cordial thanks to the referee J\"{u}rgen Blum for constructive comments.
This study was supported by a Grant-in-Aid for Scientific Research (JP18K03815) from the Japan Society for the Promotion of Science (JSPS).
We thank American Journal Experts (AJE) for English language editing.


\section*{Data Availability}

The data underlying this article will be shared on reasonable request to the corresponding author.




\bibliographystyle{mnras}
\bibliography{example} 




\clearpage

\appendix

\section{Thermal conductivity through interpebble contacts}
\label{app:k_net}

The thermal conductivity through particle contacts is modelled in our previous studies \citep{2017A&A...608L...7A, 2019Icar..324....8A, 2019PTEP.2019i3E02A}.
The thermal conductivity through interpebble contacts, $k_{\rm net}$, is given by
\begin{equation}
k_{\rm net} = 2 k_{\rm peb} \frac{r_{\rm c, peb}}{r_{\rm peb}} {f {\left( \phi_{\rm pack} \right)}},
\end{equation}
where $f$ is a dimensionless function associated with packing geometry.
\citet{2019Icar..324....8A} found that $f$ is given as a function of the filling factor, $\phi$:
\begin{equation}
{f {\left( \phi \right)}} = 0.784 \phi^{1.99} {\left( \frac{Z {\left( \phi \right)}}{2} \right)}^{0.556},
\end{equation}
where $Z$ denotes the average coordination number that also depends on $\phi$ \citep{2019Icar..324....8A, 2019PTEP.2019i3E02A}:
\begin{equation}
{Z {\left( \phi \right)}} = 2 + 9.38 \phi^{1.62}.
\end{equation}

The thermal conductivity within a pebble, $k_{\rm peb}$, is given by the sum of two terms:
\begin{equation}
k_{\rm peb} = k_{\rm net, peb} + k_{\rm rad, peb},
\end{equation}
where $k_{\rm net, peb}$ is the thermal conductivity through contacts of constituent particles and $k_{\rm rad, peb}$ is the thermal conductivity due to radiation within a pebble.

When the contact radius between constituent particles is $r_{\rm c, par}$, $k_{\rm net, peb}$ is given by
\begin{equation}
k_{\rm net, peb} = 2 k_{\rm mat} \frac{r_{\rm c, par}}{r_{\rm par}} {f {\left( \phi_{\rm peb} \right)}}.
\end{equation}
Here we note that $r_{\rm c, par}$ in the comet nucleus is nearly independent of $d$.
The external force acting on a particle, $F_{\rm par}$, is roughly given by $F_{\rm par} \sim \pi {r_{\rm par}}^{2} \sigma$ \citep[e.g.,][]{chan1973conductance, 2017AIPA....7a5310S}.
We evaluate the $x$ parameter introduced in Equation (\ref{eq:x}).
For contacts between constituent particles, we define $x_{\rm par}$ as follows:
\begin{align}
x_{\rm par} & = \frac{F_{\rm par}}{3 \pi \gamma_{\rm par} r_{\rm par}} \nonumber \\
            & \sim \frac{r_{\rm par} \sigma}{\gamma_{\rm par}} \nonumber \\
            & \sim 10^{-3} {\left( \frac{\sigma}{10^{2}~\si{Pa}} \right)},
\end{align}
and $x_{\rm par} \ll 1$ in the entire comet nucleus.
In this study, we assume that $r_{\rm c, par}$ is given by
\begin{equation}
\frac{r_{\rm c, par}}{r_{\rm par}} = {\left[ \frac{9 \pi \gamma_{\rm par} {\left( 1 - {\nu_{\rm par}}^{2} \right)}}{4 E_{\rm par} r_{\rm par}} \right]}^{1/3}.
\end{equation}

The material thermal conductivity of ice, $k_{\rm mat}$, is set to be equal to that assumed in \citet{2022MNRAS.514.3366M}.
It is known that $k_{\rm mat}$ depends on whether ice is crystalline or amorphous \citep[e.g.,][]{1980Sci...209..271K, 1992ApJ...388L..73K}.
For crystalline ice, $k_{\rm mat}$ is given by
\begin{equation}
k_{\rm mat} = 5.67 {\left( \frac{T}{100~\si{K}} \right)}^{-1}~\si{W.m^{-1}.K^{-1}},
\label{eq:k_mat_cry}
\end{equation}
while for amorphous ice, $k_{\rm mat}$ is given by
\begin{equation}
k_{\rm mat} = {\left[ 2.348 \times 10^{-1} {\left( \frac{T}{100~\si{K}} \right)} + 2.82 \times 10^{-2} \right]}~\si{W.m^{-1}.K^{-1}}.
\label{eq:k_mat_amo}
\end{equation}

When constituent particles are optically thin, \citet{2017A&A...608L...7A} found that $k_{\rm rad, peb}$ is given by
\begin{equation}
k_{\rm rad, peb} = \frac{16}{3} \sigma_{\rm SB} T^{3} l_{\rm mfp},
\end{equation}
where $\sigma_{\rm SB}$ is the Stefan--Boltzmann constant and $l_{\rm mfp}$ is the mean free path of photons.
When the Rosseland mean opacity of particles is $\kappa_{\rm R}$, $l_{\rm mfp}$ is given by
\begin{equation}
l_{\rm mfp} = \frac{1}{\kappa_{\rm R} \rho_{\rm peb}}.
\end{equation}
The density of pebbles, $\rho_{\rm peb}$, is $\rho_{\rm peb} = \rho_{\rm comet} / \phi_{\rm pack} = 887~\si{kg.m^{-3}}$.

The Rosseland mean opacity increases as $\kappa_{\rm R} \propto T^{2}$ for $T < 150~\si{K}$ \citep[e.g.,][]{1994ApJ...421..640N}:
\begin{equation}
\kappa_{\rm R} = 20 {\left( \frac{T}{100~\si{K}} \right)}^{2}~\si{m^{2}.kg^{-1}}.
\label{eq:kappa_R}
\end{equation}
For the temperature range between $50~\si{K}$ and $150~\si{K}$, we found the following relation:
\begin{equation}
r_{\rm par} \ll l_{\rm mfp} \ll r_{\rm peb},
\label{eq:optical_depth}
\end{equation}
is satisfied; in other words, mm- to cm-sized pebbles are optically thick but $1~\si{\micro m}$-sized constituent particles are optically thin in the comet nucleus.
These are consistent with the assumptions of our modelling.

\begin{figure}
\begin{center}
\includegraphics[width=\columnwidth]{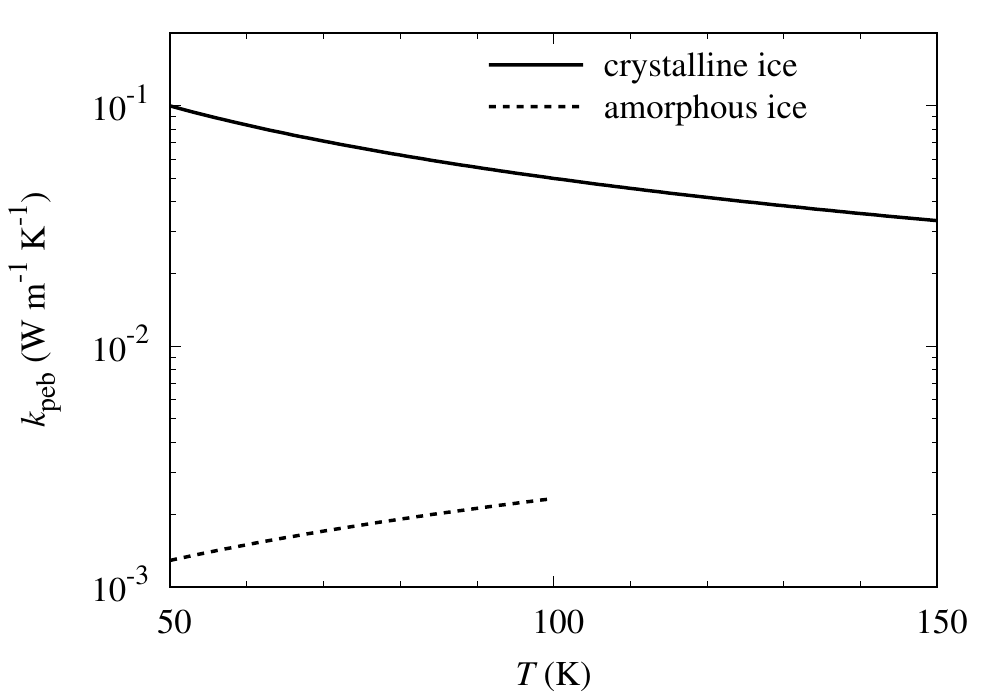}
\end{center}
\caption{
Thermal conductivity within a pebble, $k_{\rm peb}$, as a function of temperature, $T$.
We note that the temperature range of $T \le 100~\si{K}$ is considered for amorphous ice because it crystallizes at $T \simeq 100~\si{K}$.
}
\label{fig:k_peb}
\end{figure}

We found that $k_{\rm rad, peb} \sim 10^{-5}~\si{W.m^{-1}.K^{-1}}$ is orders of magnitude lower than $k_{\rm net, peb}$ and $k_{\rm peb} \simeq k_{\rm net, peb}$ in our simulations (see Figure \ref{fig:k_peb}).
We also note that the contribution of $k_{\rm rad, peb}$ is ignored in earlier studies \citep[e.g.,][]{2020MNRAS.493.3690G, 2022MNRAS.514.3366M}.

\section{Thermal conductivity due to radiation}
\label{app:k_rad}

Heat transfer due to radiation is of great importance for porous granular matter.
When the thermal conductivity due to radiation is comparable to or lower than the thermal conductivity of the constituting spheres, the nonisothermality within a sphere affects the heat transfer \citep[e.g.,][]{2020JGRE..12506100R, 2022JGRE..12707191R}.
We take into consideration this effect.

For a pebble pile, the thermal conductivity due to radiation through the void space between pebbles, $k_{\rm rad}$, is given by
\begin{equation}
k_{\rm rad} = 8 \sigma_{\rm SB} T^{3} r_{\rm peb} {{\mathscr F} {\left( \phi_{\rm pack} \right)}} {f_{\rm k} {\left( k_{\rm peb}, r_{\rm peb}, \phi_{\rm pack}, T \right)}}.
\label{eq:k_rad}
\end{equation}
Based on numerical simulations, \citet{2022JGRE..12707191R} found that the radiative exchange factor, ${\mathscr F}$, is given by
\begin{equation}
{{\mathscr F} {\left( \phi \right)}} = 0.739 + 0.629 {\left( \frac{1 - \phi}{\phi} \right)}^{1.031},
\end{equation}
and the nonisothermal correction factor, $f_{\rm k}$, is given by 
\begin{equation}
{f_{\rm k} {\left( k, r, \phi, T \right)}} = 1.007 - 0.500 \arctan{\left[ 1.351 {\left( \frac{8 \sigma_{\rm SB} r \phi T^{3}}{k} \right)}^{0.741} \right]}.
\label{eq:f_k}
\end{equation}
We note that Equation (\ref{eq:k_rad}) is applicable only when each pebble is optically thick \citep{2022JGRE..12707191R}.
We confirmed that mm- to cm-sized pebbles are indeed optically thick (see Relation (\ref{eq:optical_depth})).

\begin{figure}
\begin{center}
\includegraphics[width=\columnwidth]{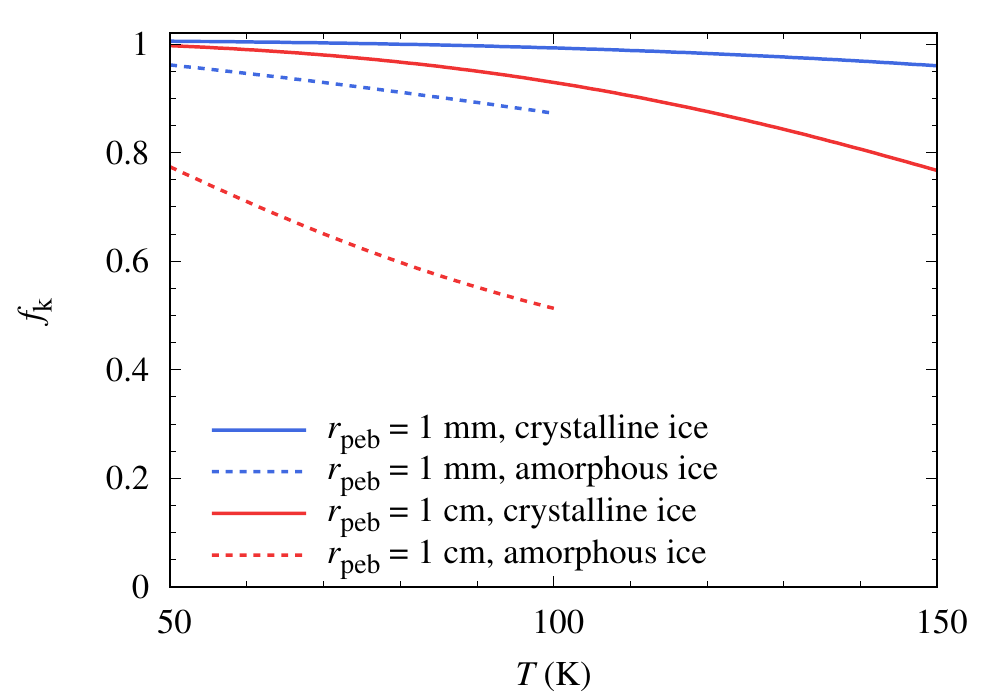}
\end{center}
\caption{
The nonisothermal correction factor, $f_{\rm k}$, as a function of the temperature, $T$ (see Equations (\ref{eq:k_rad}) and (\ref{eq:f_k})).
}
\label{fig:f_k}
\end{figure}

Figure \ref{fig:f_k} shows ${f_{\rm k} {( k_{\rm peb}, r_{\rm peb}, \phi_{\rm pack}, T )}}$ as a function of $T$.
We found that $f_{\rm k}$ for amorphous ice is lower than that for crystalline ice, and $f_{\rm k}$ decreases with increasing $r_{\rm peb}$.
We can understand these relations from Equation (\ref{eq:f_k}).
We note that the validity of Equation (\ref{eq:f_k}) has not yet been confirmed for $f_{\rm k} < 0.4$ \citep[see Figure 4 of][]{2022JGRE..12707191R}.
Future studies on the nonisothermal effect would be needed when we discuss $k_{\rm rad}$ for pebble pile comets made of larger pebbles ($r_{\rm peb} \gg 1~\si{cm}$) or higher temperature ($T \gg 150~\si{K}$).

\bsp	
\label{lastpage}
\end{document}